\documentclass[aps,pra,superscriptaddress,tightenlines,floatfix,twocolumn]{revtex4-1}
\usepackage{graphicx}
\usepackage[english]{babel}
\usepackage{amsmath}
\usepackage{amssymb}
\usepackage{tensor}

\newcommand{\vk}{\mathbf{k}}

\newcommand{\vw}{\mathbf{w}}

\newcommand{\be}{\begin{eqnarray}}
\newcommand{\ee}{\end{eqnarray}}
\newcommand{\p}{\partial}

\newcommand{\dc}{c^{\dagger}}
\newcommand{\da}{a^{\dagger}}
\newcommand{\nn}{\nonumber}

\def\ket#1{|#1\rangle}
\def\bra#1{\langle #1 |}
\def\ep#1{\langle #1 \rangle}

\begin{document}

\title{Topology, edge states, and zero-energy states of ultracold atoms in 1D optical superlattices with alternating onsite potentials or hopping coefficients}
\author{Yan He}
\affiliation{College of Physical Science and Technology, Sichuan University, Chengdu, Sichuan 610064, China}
\author{Kevin Wright}
\affiliation{Department of Physics and Astronomy, Dartmouth College, Hanover NH 03755, USA}
\author{Said Kouachi}
\affiliation{Department of Mathematics, Qassim University, Buraydah 51452, Saudi Arabia}
\author{Chih-Chun Chien}
\affiliation{School of Natural Sciences, University of California, Merced, CA 95343, USA}
\email{cchien5@ucmerced.edu}

\begin{abstract}
One-dimensional superlattices with periodic spatial modulations of onsite potentials or tunneling coefficients can exhibit a variety of properties associated with topology or symmetry. Recent developments of ring-shaped optical lattices allow a systematic study of those properties in superlattices with or without boundaries. While superlattices with additional modulating parameters are shown to have quantized topological invariants in the augmented parameter space, we also found localized or zero-energy states associated with symmetries of the Hamiltonians. Probing those states in ultracold-atoms is possible by utilizing recently proposed methods analyzing particle depletion or the local density of states. Moreover, we summarize feasible realizations of configurable optical superlattices using currently available techniques.
\end{abstract}
\maketitle

\section{Introduction}
Discoveries of materials exhibiting topological behavior have drawn intense research interest on systems with nontrivial band structures~\cite{Kane_TIRev,Zhang-TIRev,ShenTI,Asboth2016}. Quantized topological invariants can be defined and they categorize various topological systems~\cite{Chiu2016,Stanescu_book}. From the bulk-boundary correspondence, topological edge states (modes) localized at the boundary will emerge as two systems with different topological properties are connected~\cite{Kane_TIRev,Zhang-TIRev,Chiu2016,Stanescu_book}. Those topological edge states are usually also zero-energy states. On the other hand, zero-energy states can emerge if a boundary is present in a system respecting a symmetry which pairs the positive and negative energy levels. The symmetry based zero-energy states are known as Shockley states (modes)~\cite{Shockley39} and they may or may not be associated with topology~\cite{Loss12,Bello2016a}. Here we will show that, by tuning the onsite potentials or hopping coefficients of a one-dimensional (1D) superlattice, topological edge states and Shockley states can emerge in different setups.

One simple yet important 1D system elucidating topological properties of its band structure is the Su-Schrieffer-Heeger (SSH) model~\cite{Shockley39,SSH79}, which may be thought of as a superlattice with two alternating hopping coefficients. A quantized topological invariant, the winding number, can be defined for the SSH model and due to the chiral (or sublattice) symmetry, it belongs to the AIII class of topological insulators~\cite{Chiu2016}. The SSH model can support zero-energy topological edge states at the boundary, and those edge states survive even when alternating onsite potentials are present~\cite{ShenTI,Asboth2016}. It is natural to ask what happens if one considers superlattices with higher periods of alternating hopping coefficients or onsite potentials. It has been shown~\cite{shijie,Wen89} that sublattices with periodic onsite potentials and uniform hopping coefficients can be viewed as topological systems. To define quantized topological invariants, however, additional parameters need to be introduced to map the 1D superlattice to an effective 2D system~\cite{shijie}. Another generalization is to introduce alternating hopping coefficients with period three or higher. Although quantized topological invariants have not been identified for those higher-period hopping models, we found edge states and zero-energy states associated with symmetries and they serve as examples of the Shockley states.

Cold-atoms and engineered optical potentials are suitable for simulating complex many-body systems. Topological systems such as the SSH model, Haldane model, Harper model, and many others have been realized using cold-atoms~\cite{Atala2013,HaldaneHam,HarperHam}. While topological invariants in the bulk can be measured from the motion of atomic clouds, localized edge states requires sharp boundary~\cite{Goldman2013a,Mekena17}. Developments of ring-shaped lattice potentials~\cite{PaintPot2009,Amico2014} offer exciting opportunities for generating various superlattices, including the multi-period systems discussed here. To change  boundary conditions, one may insert a laser sheet and cut open the ring~\cite{WrightPRL2013}. Here we explore another type of boundary where two segments of superlattices with different periods form on the two sides of a ring lattice. While the bulk energy bands hybridize, edge states emerge at the interface and their numbers are determined by the bulk topological invariants. The optical superlattices discussed here are within current experimental capabilities and we will discuss their realizations.

To probe the topological and symmetry-based edge states, we consider two recently proposed schemes. The first one exploits the localized nature of edge states and depletes the mobile particles away from the boundary~\cite{Mekena17}. As the mobile particles are removed, the remaining density reveals the localized states originally buried in the full density profile. This method works for cold-atoms because atoms can be removed locally by laser or electron beams~\cite{Caliga2016,Barontini2013} and there is no external particle reservoir to replenish the system. Since some symmetry-based zero-energy states are not fully localized, the depletion method may not identify them. The second method probes the local density of states (LDOS) as a function of energy~\cite{Gruss17}. The symmetry-based Shockley states are zero-energy states, and they contribute to peaks in the LDOS at zero energy.

The paper is organized as follows. Section~\ref{sec:potential} discusses topological properties and edge states of 1D superlattices with multi-period potentials. Section~\ref{sec:hopping} discusses topological properties and symmetry-based states of 1D superlattices with modulating hopping coefficients. Section~\ref{sec:probe} presents possible methods for probing the edge and zero-energy states when cold-atoms are loaded into the optical superlattices discussed here. Section~\ref{sec:exp} shows possible experimental techniques for generating the superlattices and their implications. Section~\ref{sec:conclusion} concludes this work. Details of the calculations are summarized in the Appendix.

\section{1D superlattices with periodic potentials}\label{sec:potential}
We consider a family of 1D models with periodic onsite potentials described by the Hamiltonian for noninteracting fermions~\cite{shijie}
\be
H=\sum_{j=1}^N\Big[-t(\dc_j c_{j+1}+\dc_{j+1} c_j) -V\cos(2\pi\frac pq j+\theta_0) \dc_j c_j\Big].\nn\\
\label{pp1}
\ee
Here $j$ labels the sites of the 1D lattice, $t$ is the hopping coefficient, $V$ is the onsite-potential strength, $\theta_0$ is a phase angle, and $p$ and $q$ are two coprime numbers. To characterize the topology of this model, one can impose periodic boundary condition to obtain the Chern numbers or solve the eigenstates with open boundary condition to identify the edge states arising from the band topology. 

\subsection{Periodic boundary condition}\label{section:periodicbc}
We first show that topological invariants may be found in the system with periodic boundary condition, where site $N+1$ is simply site $1$.
If we treat $\theta_0$ as an independent, periodic momentum $k_y$ with $-\pi\leq k_y\leq \pi$, the model may be generalized to a 2D model
expressed in terms of the lattice coordinate $j$ and the momentum $k_y$:
\be
H&=&\sum_{j,k_y}\Big[-t(\dc_{j,k_y} c_{j+1,k_y}+\dc_{j+1,k_y} c_{j,k_y})\nn\\
& &-2V\cos(2\pi\frac pq j+k_y) \dc_{j,k_y} c_{j,k_y}\Big].
\ee
Making a Fourier transform of $k_y$ to a fictitious $y$-coordinate, we find
\be\label{eq:HPeriodicV}
H&=&\sum_{j,l}\Big[-t(\dc_{j,l} c_{j+1,l}+\dc_{j+1,l} c_{j,l})\nn\\
& &-V(e^{i2\pi\frac pq j}\dc_{j,l} c_{j,l+1}+e^{-i2\pi\frac pq j}\dc_{j,l+1} c_{j,l})\Big].
\ee
This is a 2D square lattice with a magnetic flux $p/q$ through each unit cell \cite{Hofstadter}. The magnetic unit cell is of the size $q\times 1$ when compared to the original lattice. Thus we expect the magnetic Brillouin zone is $1/q$ of the original Brillouin zone. Details of the band calculations are summarized in the Appendix.

Now we turn to some concrete examples. First, we discuss the case $p/q=1/3$, which has three bands. The onsite potential follows the pattern $(V\cos(2\pi/3+\theta_0),V\cos(4\pi/3+\theta_0),V\cos(\theta_0))$. To describe the topological properties, we discuss the Berry phase, Berry curvature, and Chern number defined as follows.
\be
&&A_{\mu}\equiv-i\bra{\psi_n}\nabla_{\mu}\ket{\psi_n},\quad
F_{xy}=\frac{\p A_y}{\p k_x}-\frac{\p A_x}{\p k_y},\nn\\
&&Ch=\frac{1}{2\pi}\int d^2k F_{xy}.\nn
\ee
Here $\ket{\psi_n}$ is the Bloch wave function of n-th band and $Ch$ is the Chern number of this band. The Chern numbers of the bands of the $p/q=1/3$ case (from the lowest to the highest) are 1, -2, 1, respectively.

\begin{figure}
\includegraphics[width=0.4\textwidth]{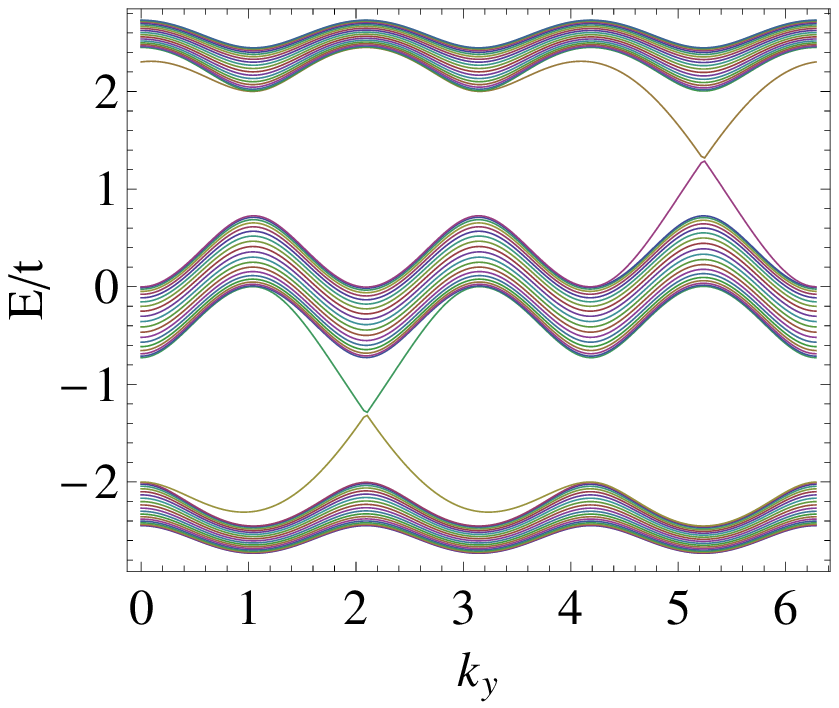}
\includegraphics[width=0.4\textwidth]{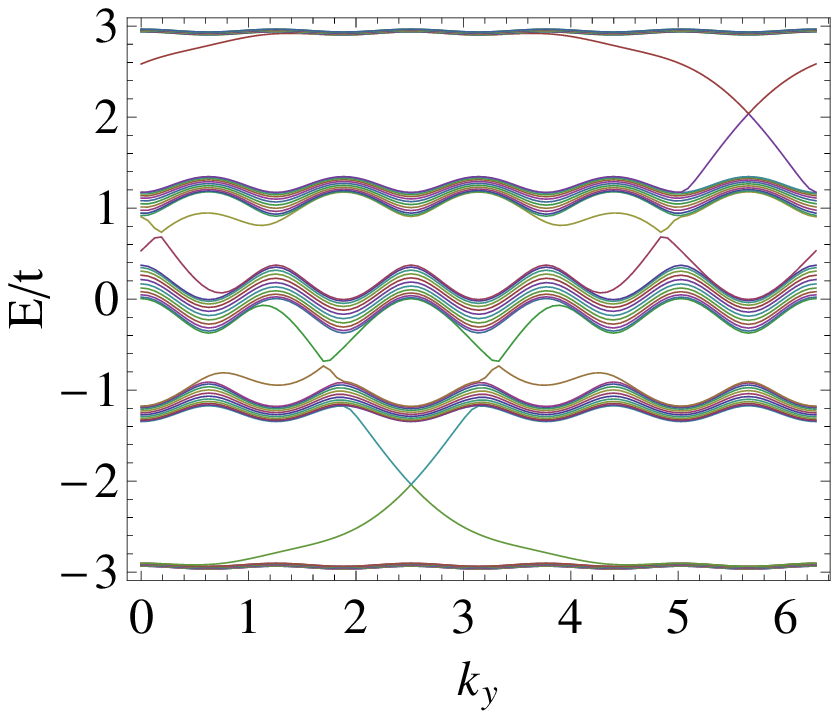}
\caption{Energy spectra as a function of $\theta_0=k_y$ for $p/q=1/3$ (left) and $p/q=1/5$ (right). The edge states are located inside the energy gaps. Here the system size is $N=60$.}
\label{q3}
\end{figure}

\begin{figure}
\includegraphics[width=0.3\textwidth]{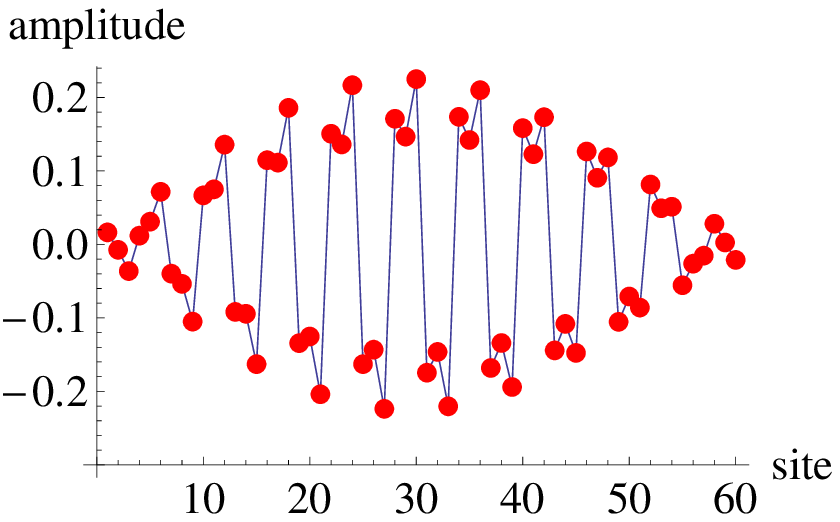}
\includegraphics[width=0.3\textwidth]{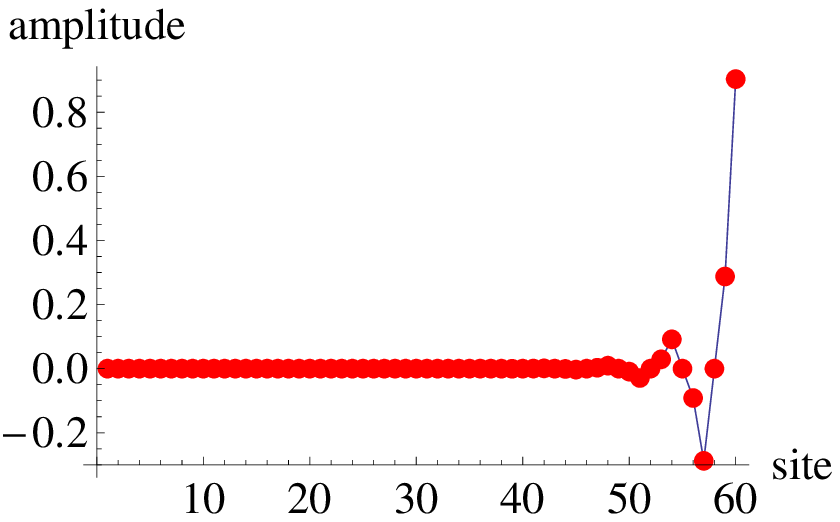}
\includegraphics[width=0.3\textwidth]{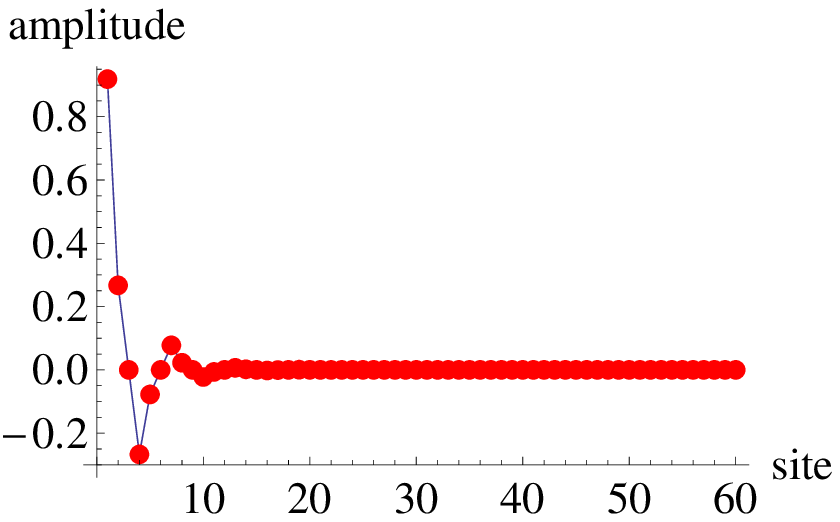}
\caption{Selected eigenstates of the bulk (top panel), right (middle panel) and left (lower panel) edge states of the $q=3$ case. Each eigen-state is normalized to $1$.}
\label{edge}
\end{figure}

Next we analyze the $p/q=1/5$ case. The Chern numbers from the lowest to highest bands are 1, 1, -4, 1, 1, respectively. The Chern number reflects the twisting of the phase of Bloch wave function. The overall phase for all bands must be trivial, therefore the sum of the Chern numbers of all bands must be zero \cite{Simon}.

\subsection{Systems with no translational invariance}
We also solve the eigenvalue problem of the Hamiltonian with open boundary condition. The resulting energy bands as a function of $k_y$ for the $q=3$ and $q=5$ cases are plotted in Figure \ref{q3}. One can clearly see in-gap states which are chiral edge states localized at the boundary. The number of chiral edge states match the Chern number from the bulk through the bulk-boundary correspondence \cite{Chiu2016}. Explicitly,
\be
n_L(E_g)-n_R(E_g)=\sum_{E_n<E_g}\mathrm{Ch}(E_n). \label{bb}
\ee
Here Ch$(E_n)$ is the Chern number of the $n$-th band. $E_g$ is one of the band gaps and $n_L(E_g)$ and $n_R(E_g)$ are the numbers of left and right moving chiral states on the left edge of the system. On the right edge, $n_L$ and $n_R$ are swapped in the formula.
In Figure \ref{edge}, we plot some selected eigenstates of this system with a chosen $\theta_0=2$. The top panel shows a typical bulk state. The middle (lower) panel shows the edge state localized on the right (left) boundary. For the $q=5$ case, the bulk-boundary correspondence can also be confirmed and the edge states can be identified.

\begin{figure}
\centerline{\includegraphics[width=0.4\textwidth]{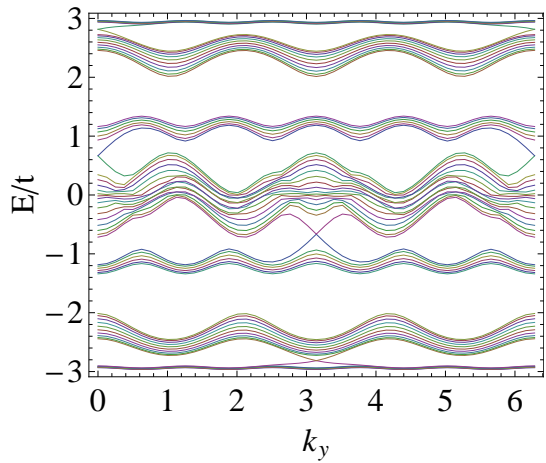}}
\caption{Energy spectrum as a function of $\theta_0=k_y$ for a ring with $N=60$ consisting of two segments with  $p/q=1/3$ and $p/q=1/5$ (with $30$ sites each).}
\label{q35}
\end{figure}

In addition to the simple open boundary condition, we present another configuration for generating boundary states by considering a system composed of segments with different periodic potentials. One example is a ring of lattices with $N$ sites, and the first $N/2$ sites have a periodic potential with period $q_1$ while the next $N/2$ sites have a periodic potential with period $q_2$. Here we consider a particular case with $N=60$ sites on a ring and $q_1=3$ and $q_2=5$ given by the following Hamiltonian
\be
&&H=\sum_{j,k_y}\Big[-t(\dc_{j,k_y} c_{j+1,k_y}+\dc_{j+1,k_y} c_{j,k_y})\Big]\nn\\
&&\quad -2V\sum_{k_y}\Big[\sum_{j=1}^{30}\cos(\frac{2\pi}3 j+k_y)\dc_{j,k_y} c_{j,k_y}\nn\\
&&\quad+\sum_{j=31}^{60}\cos(\frac{2\pi}5 j+k_y)\dc_{j,k_y} c_{j,k_y}\Big].
\ee
Although the system forms a ring, there is no lattice translational symmetry along the ring due to the two different segments. Thus, there is no exact definition of Chern number in this case. From Figure \ref{q35}, we can see the result is roughly an overlap of the period-3 and period-5 band systems. The middle band of these two system merge together to form one band, which gives rise to a total of 7 bands. When connecting the period-3 and period-5 systems to form a ring, they happen to take the position of the mirror reflection of each other. Thus, the Chern number of one system will get an extra minus sign. Therefore, we expect the Chern numbers of these 7 bands to be 1, -1, 1, -2, 1, -1, 1. One can see these numbers also match the numbers of chiral edge states through Eq. (\ref{bb}).
We caution that a more accurate definition of the Chern number may be needed for those mixed, inhomogeneous systems.

\begin{figure}
\includegraphics[width=0.3\textwidth]{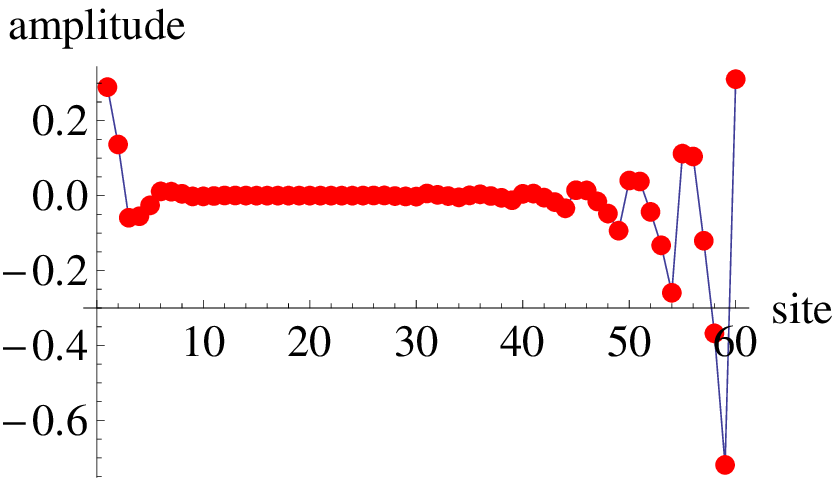}
\includegraphics[width=0.3\textwidth]{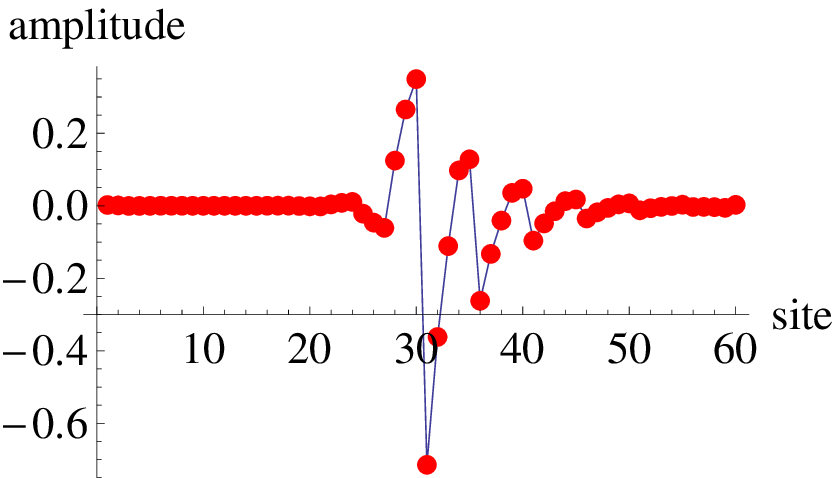}
\includegraphics[width=0.3\textwidth]{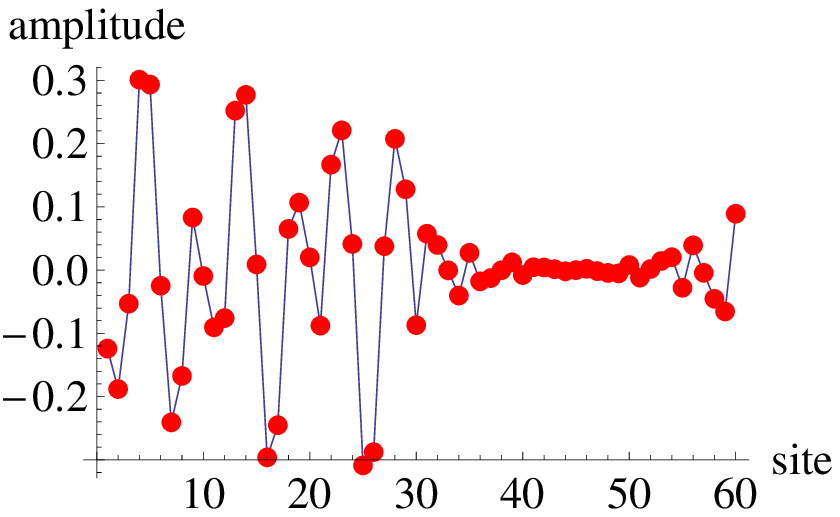}
\includegraphics[width=0.3\textwidth]{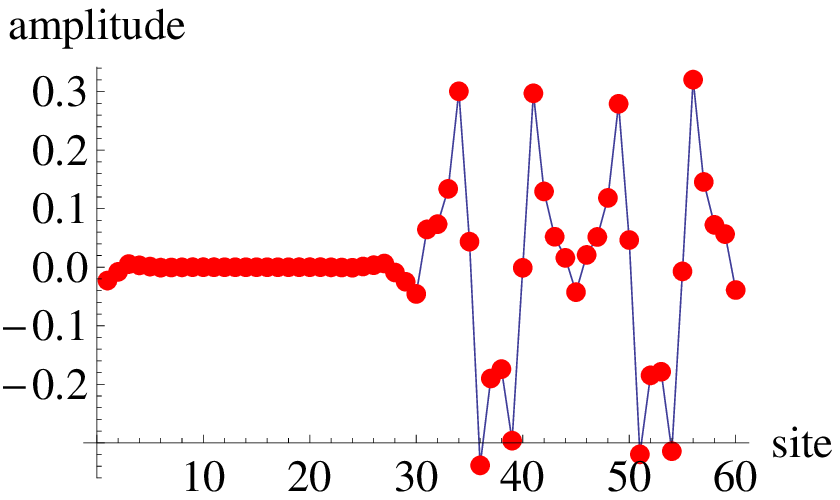}
\caption{Selected eigen-states of a ring-lattice with $L=60$ sites with the first half having a periodic-3 potential and the rest having a periodic-5 potential. The upper two panels show the two edge states between the middle band and the one below it. The two edge states are located at the boundary where the two sub-systems meet each other (sites $i=30$ and $i=60$). The lower two panels show two selected bulk states from the middle band and the band below it. Each eigen-state is normalized to $1$. Note that the middle band is dominated by the period-3 ($q_1=3$) sub-system covering the first-half part $(1\le i \le 30)$, while the band below it is dominated by the period-5 ($q_2=5$) sub-system covering the second-half part $(31\le i \le 60)$.}
\label{q35modes}
\end{figure}

The eigenstates in the mixed systems are also more complicated than their uniform counterparts. Figure~\ref{q35modes} shows some bulk states selected from the bands, and one can see that they correspond to states confined in either the first or second half with a well-defined period. Moreover, the internal boundaries where the two chains with different periodicities meet have localized edge states. We have checked other system sizes where incomplete unit cells are present and the edge states still survive. For instance, when there are $62$ sites with a periodic-3 potential on the first $31$ sites and a periodic-5 potential on the other $31$ sites, the edge states can be identified and they are localized at the two internal boundaries. The band structure, however, is quantitatively different from the $60$-site case.

\section{1D superlattices with periodic hopping coefficients}\label{sec:hopping}
The Su-Schrieffer-Heeger (SSH) model can be thought of as a 1D hopping model with two alternating hopping coefficients \cite{SSH79}. We now consider a 1D model with three alternating hopping coefficients $t_1, t_2, t_3$. To obtain the band structure, we consider three sites $i=1,2,3$ in each unit cell. In real space the Hamiltonian is
\be
H=\sum_n\Big(t_1\dc_{1,n}c_{2,n}+t_2\dc_{2,n}c_{3,n}+t_3\dc_{3,n}c_{1,n+1}+h.c.\Big).\nn\\
\label{ph}
\ee
Again the topological invariant (or localized edge states) can be found in the case with periodic (or open) boundary condition.

\subsection{Periodic boundary condition}
Although the Chern number cannot be defined in 1D, one can define a quantized winding number to characterize the topological property of the SSH model, which belongs to the AIII class in the ten-fold way classification \cite{Chiu2016}. This is different from the models with periodic onsite potentials defined in Eq.~(\ref{pp1}) which requires an extension (by introducing another effective periodic momentum) to a 2D model in order to introduce nonzero Chern numbers.

Unlike the SSH model, the period-3 hopping model of Eq.~(\ref{ph}) does not seem to have a winding number because the chiral symmetry no longer holds. It is possible, however, to extend the system to a 2D model in order to acquire nonzero Chern numbers. This can be achieved by modulating the hopping coefficients and introducing an additional dimension. For example, one can consider the periodic hopping coefficients  $(t_1\cos(k_y+\phi_1),t_2\cos(k_y+\phi_2),t_3\cos(k_y+\phi_3))$ with $-\pi\le k_y\le \pi$ and $\phi_i$, $i=1,2,3$ being some constant phases.
One has to choose suitable $t_i$ for $i=1,2,3$ to keep the gaps open. Then, the Chern numbers can be defined for this extended 2D system and for the three bands from top to bottom are $Ch=(2,\,-4,\,2)$.

\subsection{Systems  with open boundary condition}
The 1D model with periodic-3 hopping and open boundary condition can be solved in real space. By including onsite potentials of the same period, the model is given by
\be\label{eq:Hfull}
H=\sum_{i=1}^{N-1}\Big[w_i(\dc_ic_{i+1}+\dc_{i+1}c_i)+V_i\dc_ic_i\Big]+V_N\dc_Nc_N
\ee
with the hopping coefficients and onsite potentials given by
\be
w_i,~V_i=\left\{
      \begin{array}{ll}
        t_1,~V_1 & i=1;~(\hbox{mod}\,3). \\
        t_2,~V_2 & i=2;~(\hbox{mod}\,3). \\
        t_3,~V_3 & i=0;~(\hbox{mod}\,3).
      \end{array}
    \right.
\ee
If one sets $V_i=0$ for all $i$, the model reduces to one with period-3 hopping coefficients only.
By solving the eigenvalue equation $H_{ij}\psi_j=E \psi_j$, the energy bands can be obtained from numerical calculations. The details are summarized in the Appendix. In the following we will focus on three special eigenstates.

\subsection{Special eigenstates in periodic-3 hopping case}
There are three special eigenstates arising from the symmetries of the period-3 hopping model (with $V_i=0$ for all $i$) and we address them here. The first two are ``dimer states'' in the sense that inside each unit cell, two adjacent sites form a dimer with either the same or opposite amplitudes while the amplitude of the third site vanishes. Moreover, the dimer states resemble the edge states of the SSH model in the sense that the amplitude follows a power-law decay from the boundary. Those dimer states emerge when the lattice size is $N=3m+2$ for a non-negative integer $m$. Assuming $t_2<t_3$, the amplitudes of the dimer states are of the form 
\be\label{eq:edge_amp}
\mathcal{N}_d(1,\pm 1, 0, t_2/t_3, \pm(t_2/t_3),0,\cdots)
\ee
in a long lattice. The third special state is a zero-energy state, which emerges when $N=2m+1$ and $m\ge 2$. Its amplitude has the form 
\be \label{eq:zero_amp}
\mathcal{N}(1, 0, -t_1/t_2, 0, t_3/t_2, 0, -1, 0, t_1/t_2, 0, -t_3/t_2, \cdots). 
\ee
Here $\mathcal{N}_d$ and $\mathcal{N}$ are normalization factors.

We provide a simple argument to understand
the dimer states, which are shown to survive even in the presence of periodic onsite potentials. The Hamiltonian can be rewritten as
\be
&&H=H_1+H_2\nn\\
&&H_1=diag\Big\{0,\,A,\,A,\cdots,A,\,0\Big\}\\
&&H_2=diag\Big\{B,\,0,\,B,\,0\cdots,0,B\Big\}\\
&&A=\left(
        \begin{array}{ccc}
          0 & t_2 & 0 \\
          t_2 & V_3 & t_3 \\
          0 & t_3 & 0
        \end{array}
      \right),\qquad
 B=\left(
        \begin{array}{cc}
          V_1 & t_1 \\
          t_1 & V_2
        \end{array}
      \right).
\ee
Here $diag$ denotes a block diagonal matrix. $A$ has a zero mode $\psi_0=(1,0,-t_2/t_3)$ such that $A\psi_0=0$. Thus, $H_1$ also has a zero mode $\psi_1=(c_0,\,c_1\psi_0,\,c_2\psi_0,\cdots,c_m\psi_0,\,c_{m+1})$. The matrix $B$ has the following eigenvalues and eigenvectors:
\be
&&E_{1,2}=\frac{V_1+V_2}{2}\pm\sqrt{(\frac{V_1-V_2}{2})^2+t_1^2},\nn\\
&&\phi_{1,2}=(1,h\pm\sqrt{1+h^2})^T,
\ee
where $h=(V_1-V_2)/(2t_1)$. If we require $\psi_1$ to be an eigenvector of $H_2$, we find the condition
\be
&&\Big(c_i(\psi_0)_3, c_{i+1}(\psi_0)_1\Big)^T \propto \phi_{1,2},\quad i=1,\cdots,m-1,
\ee
which gives the two edge states discussed previously.

On the other hand, the zero-energy state is due to another symmetry, which arises when $V_i=0$, $N=2m+1$, and $m\ge 2$. This is actually a general results for any tridiagonal matrix with vanishing diagonal elements in odd dimensions. Construct the matrix $V=diag\{1,-1,1,-1,\cdots,-1,1\}$ and one finds $VHV=-H$. If $\psi$ is an eigenvector of $H$ with eigenvalue $E$, then $V\psi$ is an eigenvector of $H$ with eigenvalue $-E$. But the dimension of $H$ is odd, thus there must be a zero-energy state. A detailed calculation shows that the zero-energy state of the period-3 hopping model is not a localized state because its amplitude does not decay away from the boundary. Both the dimer states and the zero-energy state are from symmetries rather than topology, and they are more related to the Shockley states instead of topological edge states.

\section{Probing the edge and zero-energy states}\label{sec:probe}
In the following we discuss two possible schemes for probing the edge and zero-energy states in multi-period optical superlattices loaded with cold atoms.
\subsection{Depletion method}
\begin{figure}
\centerline{\includegraphics[width=0.9\columnwidth]{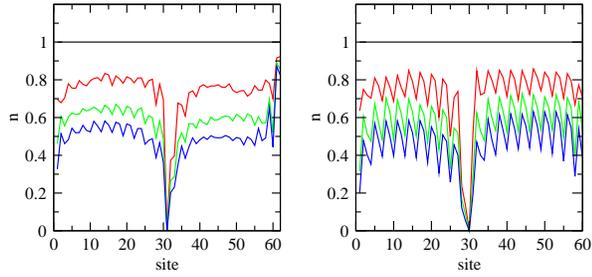}}
\caption{Evolution of density profile as particles are depleted at a selected site. The left panel shows the case of a 62-site lattice with periodic-3 hopping and onsite potential. There is an edge state on the right boundary (site $62$) of the system, and a peak there becomes visible as the overall density decreases. The curves from top to bottom correspond to $t/t_0=0,~50,~100,~150$. The right panel shows the case of a 60-site lattice with the same parameters, but in this configuration there is no edge state. Thus, the density decreases globally. Here $t_0=\hbar/t_1$, the initial filling is one particle per site, and $t_2/t_1=1.4$, $t_3/t_1=0.7,V_1/t_1=0.1$, $V_2/t_1=0.2$, $V_3/t_2=0.3$. The depletion site is at $i=31$ ($i=30$) for the first (second) case.}
\label{dep}
\end{figure}

A method for observing localized states in cold-atom systems by depleting the mobile atoms away from the localized regions has been discussed in Ref.~\cite{Mekena17}. Here we use the depletion method to probe the dimer states in the periodic-hopping model with open boundary condition. Figure \ref{dep} shows the time evolution of density profile when a localized depletion beam removes the atoms on a selected site away from the boundary. We take a generic case with $t_2/t_1=1.4$, $t_3/t_1=0.7$, $V_1/t_1=0.1$, $V_2/t_1=0.2$, $V_3/t_1=0.3$, and use the time unit $t_0=\hbar/t_1$. To contrast the edge-state effect, we compare two cases. The first one has $L=62$ sites and from the previous analysis it supports an edge state (on the right boundary for the selected parameter), and the second one has $L=60$ sites and does not have any edge state. The initial state has one fermion per site and the evolution of the correlation matrix is given by
\be
\frac{d\ep{\dc_ic_j}}{dt}&=&-i\ep{[\dc_ic_j,\,H]}.
\ee
The density at site $j$ is given by $n_j=\ep{\dc_jc_j}$. The initial condition corresponds to a completely filled lattice and its correlation matrix is $\ep{\dc_ic_j}(t=0)=\delta_{ij}$. The depletion on a selected site $m$ is simulated by setting  $\ep{\dc_mc_i}=\ep{\dc_ic_m}=0$ every $t_0/2$. The results are insensitive to the depletion rate, but an extremely high depletion rate may resemble a hard-wall potential and does not lead to efficient depletion~\cite{Mekena17}.

We demonstrate the depletion method by considering a lattice of $62$ sites with open boundary condition and deplete the atoms in the middle, $m=31$ site. As the mobile atoms are depleted from the lattice, one can see a saw-tooth pattern emerged at the right boundary, and the maximal density near the right edge stops decreasing as time evolves. Therefore, the localized edge state surfaces as the mobile atoms are removed. In contrast, for a nearly identical system with $L=60$ sites which does not support any edge state, a depletion beam at the $m=30$ site lowers the density profile in a global fashion and there is no special feature emerging at the boundary.

\begin{figure}
\centerline{\includegraphics[width=0.4\textwidth]{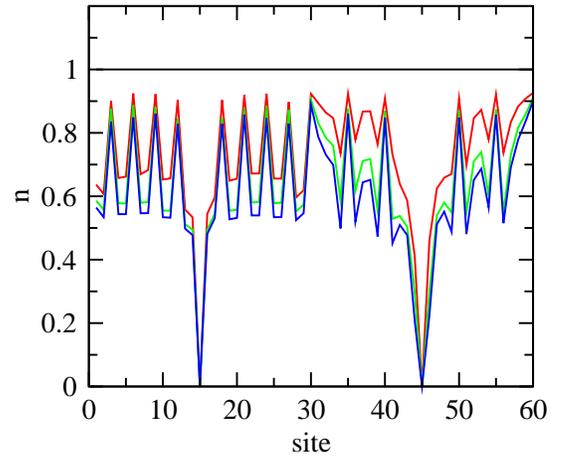}}
\caption{Evolution of density profile of a ring lattice composed of two segments with different structures. The curves from top to bottom show the density profiles at $t/t_0=0,50, 100,150$ and the system is a 60-sites lattice. The first half labeled as $i=1,\cdots,30$ has period-3 onsite potentials and the other half ($i=31,\cdots,60$) has period-5 onsite potentials. The hopping coefficient is uniform. The depletion beam is placed at site $15$ and site $45$. There are edge states appearing around sites 30 and 60, which are the interfaces between the two sub-systems, but the density profile also shows peaks corresponding to the potential period on each side.}
\label{dep-p35}
\end{figure}

In addition to using an open chain for analyzing the boundary effects, we also discuss possibilities of generating internal boundaries by connecting two superlattices with different topological properties into a ring. Figure~\ref{dep-p35} shows the time-evolution of density profile of a ring of $N=60$ sites with half of it having period-3 on-site potentials and the other half having period-5 on-site potentials. Two depletion sites, one in the middle of the period-3 segment and the other in the middle of the period-5 segment, remove particles from the system. The equations of motion for the correlation matrix follow the Heisenberg equation. We consider uniform hopping coefficients $t_i=t_1$ for all $i$,
$V_i/t_1=2\cos(\frac{2\pi i}{3}+k_y)$ for $i=1,\cdots,30$ and
$V_i/t_1=2\cos(\frac{2\pi i}{5}+k_y)$ for $i=31,\cdots,60$ with $k_y=3.1$
The initial condition is again one fermion per site, $\ep{\dc_ic_j}(t=0)=\delta_{ij}$, and the depletion is simulated by setting $\ep{\dc_mc_i}=\ep{\dc_ic_m}=0$ every $t_0/2$, where the depletion sites are chosen at $m=15,45$ for the $60$-site lattice.

For this composite system, the wavefunction is more complicated and Figure~\ref{dep-p35} shows that instead of only exhibiting localized states, the evolution of density profile also shows oscillating density peaks reflecting the periods of the underlying onsite potentials. Therefore, it is more challenging to distinguish bulk and boundary properties in a composite system with segments possessing different topological properties.

\subsection{Local Density Of States}
On the other hand, the zero-energy state in the model with period-3 hopping coefficients is not a localized one because its amplitude takes a periodic form throughout the bulk. Thus, the depletion method does not reveal its presence. Here we propose a possible probe for detecting the zero-energy state based on a recently developed method called the energy-resolved atomic spectroscopy (ERASP)~\cite{Gruss17}, which is capable of mapping out the local density of states (LDOS) of a given many-body system. The ERASP consists of a narrow-bandwidth noninteracting lattice connecting to a selected point of the system which one wants to measure, and the steady-state current siphoned from the system is proportional to the LDOS. The LDOS is defined as 
\begin{equation}\label{eq:LDOS}
D(j,E)=\sum_{n} |\ep{j|\phi_n}|^2\delta(E-E_n),
\end{equation}
where $\ket{\phi_n}$ is the n-th eigenstate with energy $E_n$ and $\ket{j}$ is the state localized at site $j$.

\begin{figure}[t]
\centerline{\includegraphics[width=0.3\textwidth]{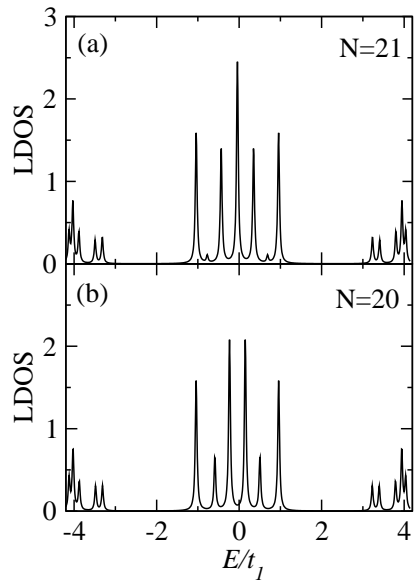}}
\caption{The LDOS (in units of $1/t_1$) on a selected site (site-$5$) of a period-3 hopping model with open boundary condition and $t_2/t_1=2$, $t_3/t_1=3$, and $V_i=0$ for all $i$. The upper panel shows the case with $N=21$ sites supporting a zero-energy state. The lower panel shows the case with $N=20$ sites and no zero-energy state.}
\label{LDOS}
\end{figure}

Figure \ref{LDOS} shows the LDOS of a model with only periodic hopping. Since the zero-energy state spreads into the bulk, its amplitude decreases as the system size increases because the state has to be properly normalized. Thus, it is easier to observe effects from the zero-energy state if the system size is small. 

Here we illustrate the LDOS signature of the zero-energy state by a period-3 hopping system with open boundary condition and $t_2/t_1=2$, $t_3/t_1=3$, and $V_i=0$ for all $i$. To contrast the zero-energy state, we consider two cases, one with $N=21$ sites supporting a zero-energy state and one with $N=20$ sites without any zero-energy state. In our numerical calculations of the LDOS, the delta function is replaced by a Lorentzian function with width $0.05t_1$.
In both cases shown in Figure \ref{LDOS}, we take $\ket{j=5}$ to probe the LDOS on site $5$. The selection of which site to probe is guided by the amplitude form of the zero-energy state shown in Eq.~\eqref{eq:zero_amp}. Apparently there is a peak at zero energy in the lattice with an odd-number of sites reflecting the existence of a zero-energy state. There is no zero-energy peak in the lattice with an even number of sites. As the system size increases, there are more peaks around the zero-energy and it is harder to resolve the zero-energy state.

\section{Experimental Implementation}\label{sec:exp}
The results presented above pertain to discrete lattice models in the tight-binding limit. Experimental implementation of these models will require generating continuous lattice potentials that suitably approximate Hamiltonians such as Eqns.~\eqref{pp1} and~\eqref{ph}. Optical dipole traps can be used for this purpose, with appropriately structured light beams generated by amplitude or phase techniques for spatial light modulation (SLM)\cite{McGloin2003, Lee2014}. The relationship between the SLM pattern and the dipole potential is generally not simple or linear when generating features near the resolution limit of the optical system, and ultimately close to the wavelength of the light used for the dipole trap. These challenges can largely be solved by numerical techniques and good optical design, but it is not our intention to address these issues thoroughly here. The examples given below are intended to demonstrate that the proposed experiments are feasible with current optical techniques. 

\begin{figure}[t]
\centerline{\includegraphics[width=0.45\textwidth]{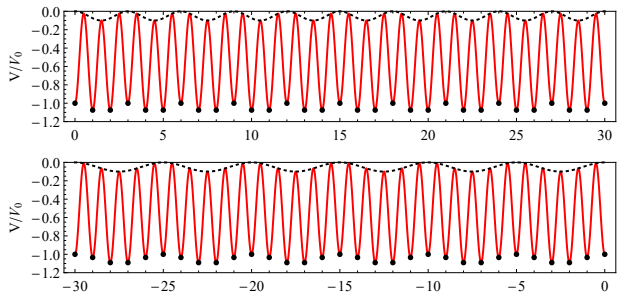}}
\caption{Continuous 1D lattice potentials of the form given by equation~\eqref{var_contrast}. Top: $p=1$, $q=3$, $\theta_0 = 0$, and $\eta=0.1 V_0$. Bottom: $p=1$, $q=5$, $\theta_0 = 0$, and $\eta=0.1 V_0$ (bottom). The horizontal axis is the scaled lattice coordinate $x/a$. The dotted black line highlights the modulation of the lattice maximum, and and the black circles highlight the modulation of the lattice minimum, which approximately realizes the \emph{discrete} lattice potential in the Hamiltonian of Eq.~\eqref{pp1}.}
\label{latticeplots_var_onsite}
\end{figure}

In ring lattices of the type introduced in Section~\ref{section:periodicbc}, the on-site energy is modulated periodically, and the tunneling is kept (approximately) constant. To achieve this in a continuous potential, the lattice contrast must be modulated along with the potential depth at each site. The continuous 1D potential around the ring should generally resemble the following:
\begin{equation}
V(x)=V_0 \left[ \cos^2\left(\pi\frac{x}{a}\right) + \eta\cos^2\left(\pi\frac{p}{q}\frac{x}{a}  +\theta_0.\right)\right]\label{var_contrast},
\end{equation}
where $x$ is the azimuthal coordinate, $a$ is the lattice constant, $V_0$ is the depth of the un-modulated lattice, and $\eta$ is the fractional modulation depth. Figure~\ref{latticeplots_var_onsite} shows two examples of smooth 1D lattice potentials of this form.

The key feature of these modulated lattice potentials is that the peak-to-valley amplitude is approximately constant, while the minima vary periodically. There are several viable techniques for creating 3D optical potentials with this type of modulated 1D profile. Here we  specifically highlight the use of a digital micromirror device (DMD) as an amplitude spatial light modulator in a high-resolution laser projection system.

\begin{figure}[t]
\centerline{\includegraphics[width=0.5\textwidth]{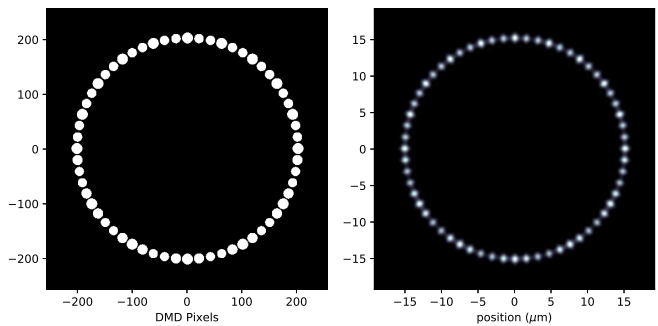}}
\caption{Calculated 60-site ring lattice potential produced by imprinting the binary image shown at left on a digital micromirror device (DMD). White is ``on'', black is ``off''. The image at right is a scalar diffraction calculation of the intensity profile at the focal plane of an optical system projecting the DMD pattern onto atoms trapped in a horizontally oriented ``sheet'' potential. The diffraction calculation assumes $\lambda$ = 780 nm, a (de)magnification factor of 100 and an image numerical aperture of 0.55. The upper half of the ring lattice has a period-3 site-depth modulation, and the lower half is period-5. The site spacing is 1.6 $\mu$m}
\label{3-5_lat}
\end{figure}

Some of the possibilities afforded by using a digital micromirror device with megapixel resolution can be illustrated by scalar diffraction calculations such as the one in Fig.~\ref{3-5_lat}. In that example, an initially uniform-intensity beam is incident on the central 512$\times$512 pixel region of a DMD, with a binary ring-lattice pattern applied as shown at left. The pixel size for a typical commercially available DMD is 7.5 $\mu$m, and in this example each of the 60 sites is created by an ``on'' region with a radius of 10 pixels on average. Imaging the plane of the DMD into an experimental chamber with a suitably large degree of demagnification reproduces this pattern, spatially filtered by the optical transfer function of the imaging system. The minimum realistic lattice spacing is therefore a function of the laser wavelength and image numerical aperture (NA).  With a wavelength of $\lambda$=780 nm and an image NA of 0.55, lattice spacings of $a$ = 1.6 $\mu$m are possible.

Modulation of the ring lattice can be achieved by several means. The simplest is to change the area of the DMD associated with each site, which alters the depth of that site. The control is not linear when the lattice spacing is close to the resolution limit, because of the convolution of the DMD pattern with the point spread function. It is also possible to control tunneling rates somewhat independently of the site depth by controlling shape and spacing of each bright ``site'' on the DMD. Numerical analysis of the 3D potential is required for deterministic control of the tunneling rates, in general.

\begin{figure}[t]
\centerline{\includegraphics[width=0.5\textwidth]{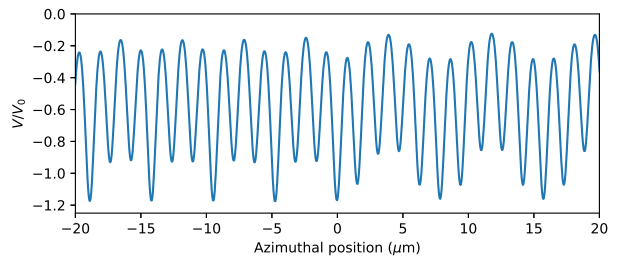}}
\caption{Depth of the ring lattice shown at left in Fig.~\ref{3-5_lat} as a function of azimuthal distance from the right-side boundary between the period-3 and period-5 regions. Counter-clockwise is positive.}
\label{3-5_lat-depth}
\end{figure}

The example lattice depicted in Fig.~\ref{3-5_lat} has two regions of different periodicity: the top half is period-3, and the bottom half is period-5. The modulation is more apparent in Figure ~\ref{3-5_lat-depth} , which is a plot of the depth of the lattice potential in Fig.~\ref{3-5_lat} as a function of azimuthal position around the ring. This example has an integer number of unit cells in each of the two regions, but we note that there is no obstacle to creating non-integer unit cells, which would lead to the effects discussed in section~\ref{section:periodicbc}. Furthermore, current DMD technology allows switching speeds as fast as 15 kHz, which will enable dynamical experiments where the ring is split at one or more locations simply by moving the sites. Lattice beams similar to the one displayed in Fig~\ref{3-5_lat} have already been demonstrated at a more modest numerical aperture of 0.32~\cite{realstuff}.

The results shown in Figs.~\ref{latticeplots_var_onsite}-\ref{3-5_lat-depth} show lattice modulations that are larger than would likely be used in an experiment, for clarity. Achieving sufficient control of the potential in the presence of noise and laser speckle will be challenging, Some, but not all of the techniques proposed above might require further development of techniques for loading the ring lattice, and reducing the temperature of the system sufficiently. Fortunately, many of the interesting features of these systems above are robust enough to tolerate realistic levels of imprecision in control of tunneling rates and on-site energies. Furthermore, topological invariants, such as the Chern number or winding number, are well defined for the eigenstates of noninteracting systems. At finite temperatures, the system is in a mixed state and the definitions of topological properties may need to be modified~\cite{Viyuela14,Bardyn17}. However, the edge states and zero-energy states discussed here are protected by symmetries of the Hamiltonians, so they still exist at the boundary and temperature only changes their relative occupation. Therefore, the depletion method and the ERASP should still capture the features presented here although the signal can be reduced due to the underlying thermal distribution. In the presence of interactions, topological properties can be different from their noninteracting counterparts~\cite{Chiu2016,Stanescu_book}. In this work we focus on noninteracting fermions to clarify their topological or symmetrical properties in ring-shape superlattices. Given the broad tunability of interactions in cold-atoms systems~\cite{Pethick_book,Stanescu_book}, future studies of interacting systems in multi-period superlattices will lead to more interesting phenomena.

\section{Conclusion}\label{sec:conclusion}
By extending the 1D periodic potential or periodic hopping models with additional periodic parameters, those models can be generalized to 2D topological systems and the Chern numbers can be defined for the extended 2D systems. Localized topological edge states are shown to emerge at the boundary in compliance with the bulk-boundary correspondence. 

On the other hand, the 1D periodic hopping model can possess localized states and zero-energy states associated with symmetries but not topology of the Hamiltonian. By considering recently developed methods for generating ring-shaped optical superlattices and probing localized states or the LDOS in cold-atom systems, measurement and characterization of interesting states in those systems are experimentally feasible. Composite systems with segments of different topological systems can exhibit internal boundaries and may inspire future investigations.

Acknowledgement -- Y. H. is supported by MOST of China under the grant No. 2017YFB0405700.

\begin{widetext}
\appendix
\section{Details of band calculations}
After obtaining Eq.~\eqref{eq:HPeriodicV}, we can obtain the band structure by a Fourier transform of the Hamiltonian to momentum space and get
\be
H=\sum_{k_x,k_y}\Big[-2t\cos k_x \dc_{\vk}c_{\vk}
-V(e^{-ik_y}\dc_{\vk-\vw} c_{\vk}+e^{ik_y}\dc_{\vk+\vw} c_{\vk})\Big]
\ee
for $-\pi\leq k_x,k_y\leq\pi$ and $\vw=(2\pi p/q,\,0)$.
To adapt to the magnetic Brillouin zone, we define $q$ species of fermions $a_{n,\vk}=c_{\vk+n\vw}$ for $-\pi/q\leq k_y\leq \pi/q$ and rewrite the Hamiltonian as
\be
H=\sum_{k_x,k_y}\sum_{n=1}^q\Big[-2t\cos(k_x+2\pi\frac pq n) \da_{n,\vk}a_{n,\vk}
-V(e^{-ik_y}\da_{n,\vk} a_{n+1,\vk}+e^{ik_y}\da_{n+1,\vk} a_{n,\vk})\Big].
\ee
In the Bloch wavefunction basis, the Hamiltonian is a $q$ by $q$ matrix:
\be
H=\left(
    \begin{array}{cccc}
      -2t\cos(k_x+w) & -Ve^{-ik_y} & \cdots & -Ve^{ik_y}\\
      -Ve^{ik_y} & -2t\cos(k_x+2w) &  & \vdots \\
      \vdots &  & \ddots & -Ve^{-ik_y} \\
      -Ve^{-ik_y} & \cdots & -Ve^{ik_y} & -2t\cos(k_x+qw) \\
    \end{array}
  \right)
\ee
with $w=2\pi p/q$. The eigenvalues and eigenvectors can then be found.

The energy spectrum of the Hamiltonian \eqref{eq:Hfull} with open boundary condition can be obtained by solving the corresponding eigenvalue problem.
For $N=3m$, the time-independent Schrodinger equation in the matrix form gives
\be
\left(
  \begin{array}{cccccc}
    u_1 & t_1 & 0 & 0 & \cdots & 0 \\
    t_1 & u_2 & t_2 & 0 & \cdots & 0 \\
    0 & t_2 & u_3 & t_3 & \cdots & 0 \\
    0 & 0 & \ddots & \ddots & \ddots & \vdots \\
    \vdots & \vdots & \ddots & \ddots & u_2 & t_2 \\
    0 & 0 & \cdots & 0 & t_2 & u_3
  \end{array}
\right)
\left(
  \begin{array}{c}
    x_1 \\
    x_2 \\
    x_3 \\
    \vdots \\
    x_{N-1} \\
    x_N
  \end{array}
\right)=0.
\label{ev}
\ee
Here $u_i=V_i-E$. From the second to the $(N-1)$-th rows, we find the following recursive relations
\be
t_1x_{3i+1}+u_2 x_{3i+2}+t_2x_{3i+3}=0,\\
t_2x_{3i+2}+u_3 x_{3i+3}+t_3x_{3i+1}=0,\\
t_3x_{3i+3}+u_1 x_{3i+1}+t_1x_{3i+2}=0.
\ee
From the above relations, we find
\be
t_1t_2t_3x_{3i+r}+\Big[u_1u_2u_3-(u_1t_2^2+u_2t_3^2+u_3t_1^2)\Big]x_{3(i+1)+r}+
t_1t_2t_3x_{3(i+2)+r}=0
\label{x33}
\ee
for $r=1,2,3$. Let
\be
u_1u_2u_3-(u_1t_2^2+u_2t_3^2+u_3t_1^2)+2t_1t_2t_3\cos\theta=0.
\label{u-th2}
\ee
Then we find the following solution to Eq.~(\ref{x33}):
\be
x_{3n+j}=C_j e^{in\theta}+D_j e^{-in\theta},~j=1,2,3.
\ee
The coefficients $C_j$ and $D_j$  are determined by the boundary condition provided by the first five and last rows of Eq.~(\ref{ev}). The resulting equations are
\be
\left(
  \begin{array}{cccccc}
    u_1 & u_1 & t_1 & t_1 & 0 & 0 \\
    t_1 & t_1 & u_2 & u_2 & t_2 & t_2 \\
    t_3a & t_3b & t_2 & t_2 & u_3a & u_3b \\
    u_1a & u_1b & t_1a & t_1b & t_3 & t_3 \\
    t_1a & t_1b & u_2a & u_2b & t_2a & t_2b \\
    0 & 0 & t_2a^{m-1} & t_2b^{m-1} & u_3a^{m-1} & u_3b^{m-1}
  \end{array}
\right)
\left(
  \begin{array}{c}
    C_1 \\
    D_1 \\
    C_2 \\
    D_2 \\
    C_3 \\
    D_3 \\
  \end{array}
\right)=0,
\ee
where $a=e^{i\theta}$ and $b=e^{-i\theta}$. The condition for a non-zero solution is that the determinant of the above coefficient matrix is zero, which gives the following relation:
\be
&&t_3f_1(\theta)u_2-t_1t_2f_2(\theta)=0,\\
&&f_1(\theta)=\sin(m+2)\theta-2\sin m\theta+\sin(m-2)\theta,\\
&&f_2(\theta)=\sin(m+3)\theta-2\sin (m+1)\theta+\sin(m-1)\theta.
\ee
Solving the above equations with Eq.~(\ref{u-th2}) and $u_i=V_i-E$ determines $E$ and $\theta$ together.

Similarly, for $N=3m+1$, the last row eigenvalue equation is $t_3x_{N-1}+u_1x_N=0$, which becomes the following boundary condition
\be
u_1a^mC_1+u_1b^mD_1+t_3a^{m-1}C_3+t_3b^{m-1}D_3=0.
\ee
Setting the determinant of the above coefficient matrix to zero gives the following relation
\be
t_2f_2(\theta)u_1-t_1t_3f_1(\theta)=0.
\ee

Similarly, for $N=3m+2$, the last-row eigenvalue equation is $t_1x_{N-1}+u_2x_N=0$, which becomes the following boundary condition
\be
u_2a^mC_2+u_2b^mD_2+t_1a^{m}C_1+t_1b^{m}D_1=0.
\ee
Again, setting the determinant of the above coefficient matrix to zero gives the following relation
\be
f_2(\theta)(u_1u_2-t_1^2)=0.
\ee
For the $N=3m+2$ case, there are two special eigenvalues from $u_1u_2=t_1^2$. They are
\be
E_1=\frac{V_1+V_2}{2}-\sqrt{(\frac{V_1-V_2}{2})^2+t_1^2},\qquad
E_2=\frac{V_1+V_2}{2}+\sqrt{(\frac{V_1-V_2}{2})^2+t_1^2}.
\ee
The corresponding eigenvectors are $v_{1,2}=\mathcal{N}_d (x_1,\cdots,x_N)$ with
\be
&&x_{3i+1}=\Big(-\frac{t_2}{t_3}\Big)^i(\sqrt{1+h^2}\pm h)^i,\quad \mbox{for},\,i=0,\cdots,m.\\
&&x_{3i+2}=\pm\Big(-\frac{t_2}{t_3}\Big)^i(\sqrt{1+h^2}\pm h)^{i+1},\quad \mbox{for},\,i=0,\cdots,m.\\
&&x_{3i+3}=0,\quad \mbox{for},\,i=0,\cdots,m-1.
\ee
Here $\mathcal{N}_d$ is a normalization factor and $h=(V_1-V_2)/(2t_1)$. If $t_2<t_3$, the above two states are edge states localized at the boundary. There is no such edge state for the $N=3m$ and $N=3m+1$ cases because the symmetry is not respected.

\end{widetext}

\bibliographystyle{apsrev}

\end{document}